\documentclass[reprint,noshowpacs,noshowkeys,prd,balancelastpage,nofootinbib]{revtex4}
\usepackage{amsfonts}
\usepackage{mdframed}
\usepackage{amssymb}
\usepackage{footnote}
\usepackage{amsmath}
\usepackage{graphicx}
\usepackage{float}
\usepackage[font={footnotesize,it}]{caption}
\usepackage[utf8]{inputenc}
\usepackage{natbib}
\usepackage{tikz}
\usepackage{tikz-3dplot}
\usepackage[colorlinks=true,
            linkcolor=purple,
            urlcolor=purple,
            citecolor=blue]{hyperref}

\begin{document}

\title{Schwarzschild-Levi-Civita black hole }
\author{S. Habib Mazharimousavi}
\email{habib.mazhari@emu.edu.tr}
\affiliation{Department of Physics, Faculty of Arts and Sciences, Eastern Mediterranean
University, Famagusta, North Cyprus via Mersin 10, Turkiye}

\begin{abstract}
Inspired by the geometry of the Ernst black hole which interpolates the
static spherically symmetric Schwarzschild black hole and the static axially
symmetric magnetic Melvin spacetime, in this paper, we aim to present a
solution to Einstein's field equation in vacuum to interpolate the
Schwarzschild black hole and the Levi-Civita spacetime. Hence, we shall call
the solution the Schwarzschild-Levi-Civita black hole. Effectively the
solution is a two-parameter non-asymptotically flat black hole that is
static, axially symmetric, and singular in its axis of symmetry.
\end{abstract}

\date{\today }
\maketitle

\section{Introduction}

One of the axially symmetric black hole spacetime is the so-called Ernst
black hole \cite{Ernst} with the line element%
\begin{equation}
ds^{2}=\Lambda ^{2}\left( -\left( 1-\frac{2m}{r}\right) dt^{2}+\frac{dr^{2}}{%
1-\frac{2m}{r}}+r^{2}d\theta ^{2}\right) +\frac{r^{2}\sin ^{2}\theta d\phi
^{2}}{\Lambda ^{2}},  \label{I1}
\end{equation}%
in which 
\begin{equation}
\Lambda =1+\frac{1}{4}B^{2}r^{2}\sin ^{2}\theta .  \label{I2}
\end{equation}%
The black hole (\ref{I1}) is a 2-parameter static axially symmetric solution
of Einstein-Maxwell equations with an external magnetic field $B$ in the
polar direction. The solution reduces into the Schwarzschild black hole in
the absence of the magnetic field i.e., $B=0,$ and coincides with Melvin
magnetic universe with $m=0$. For these two extreme limits usually the
solution (\ref{I1}) is also referred to as the Schwarzschild black hole
immersed into a magnetic field or the magnetized Schwarzschild black hole.
The Ernst black hole has been widely studied in the literature. One of the
very recent studies was done by Senjaya in \cite{Senjaya} where the exact
massless scalar quasibound state in the Ernst black hole background is
presented. In \cite{Becar} Becar et al. studied the quasinormal modes of a
charged scalar field in the same background spacetime. Gibbons et al. in 
\cite{Gibbons} reconsidered the original rotating Ernst black hole and
showed that it asymptotically doesn't mimic the Melvin metric unless the
electric charge of the black hole and the external magnetic field satisfy $%
Q=jB\left( 1+\frac{1}{4}j^{2}B^{4}\right) $ in which $Q$ and $B$ are the
electric charge of the black hole and the external magnetic field,
respectively, and $j$ is the angular momentum of the Kerr-Newman black hole
which was used as a seed solution in the Ernst procedure \cite{Wild}.
Furthermore, there are significant details on Ernst's procedure and the
rotating Ernst solution with the electromagnetic fields in \cite{Gibbons}
which makes it essential to study Ernst's black hole.

In this current study, we present a new black hole solution that is similar
to the Ernst static black hole (\ref{I1}) but instead of interpolating the
Schwarzschild black hole and Melvin magnetic universe, it interpolates the
Schwarzschild black hole and the Levi-Civita spacetime. Both the
Schwarzschild black hole and the Levi-Civita spacetime \cite{Levi} are
solutions to Einstein's equations in vacuum. Hence, unlike the Ernst black
hole, the Schwarzschild-Levi-Civita black hole - this is what we shall call
our solution hereafter - is also a solution of Einstein's field equation in
vacuum. In general relativity, except for the Schwarzschild black hole and
the Levi-Civita spacetime which are static, there exists another solution to
Einstein's equations in vacuum which is known as $\gamma $-metric or
Zipoy-Voorhees spacetime \cite{Z,V}. All these solutions are very well-known
in the literature due to their exactness and clarity which make them
practical. Having an exact solution to Einstein's field equation in its own
right is very important and deserves to be searched for. In the solution
presented in this paper, we observe not only its exactness but also its
simplicity and clarity, which make the study interesting in its own right.

\section{Field equations and the black hole solution}

Inspired by the Ernst black hole solution (\ref{I1}), we consider the
Schwarzschild-Levi-Civita black hole to be described by the line element 
\begin{equation}
ds^{2}=\Lambda ^{2}\left( -f\left( r\right) dt^{2}+\frac{dr^{2}}{f\left(
r\right) }+r^{2}d\theta ^{2}\right) +\Lambda ^{-2}r^{2}\sin ^{2}\theta d\phi
^{2},  \label{1}
\end{equation}%
in which in (\ref{1}) $\Lambda =\Lambda \left( r,\theta \right) $. We aim to
solve Einstein's vacuum field equations i.e., $G_{\mu }^{\nu }=0$.
Concerning the static line element (\ref{1}) one explicitly obtains the
field equations given by 
\begin{equation}
G_{t}^{t}=\frac{f\Lambda _{,r}^{2}}{\Lambda ^{4}}-\frac{2f\Lambda _{,r}}{%
r\Lambda ^{3}}+\frac{f^{\prime }}{r\Lambda ^{2}}+\frac{\Lambda _{,\theta
}^{2}}{r^{2}\Lambda ^{4}}-\frac{2\Lambda _{,\theta }}{r^{2}\Lambda ^{3}}\cot
\theta -\frac{1-f}{r^{2}\Lambda ^{2}}=0,  \label{2}
\end{equation}%
\begin{equation}
G_{r}^{r}=-\frac{f\Lambda _{,r}^{2}}{\Lambda ^{4}}+\frac{2f\Lambda _{,r}}{%
r\Lambda ^{3}}+\frac{f^{\prime }}{r\Lambda ^{2}}+\frac{\Lambda _{,\theta
}^{2}}{r^{2}\Lambda ^{4}}-\frac{2\Lambda _{,\theta }}{r^{2}\Lambda ^{3}}\cot
\theta -\frac{1-f}{r^{2}\Lambda ^{2}}=0,  \label{3}
\end{equation}%
\begin{equation}
G_{r}^{\theta }=\frac{2\Lambda _{,r}\Lambda _{,\theta }}{r^{2}\Lambda ^{4}}+%
\frac{2\Lambda _{,r}}{r^{2}\Lambda ^{3}}\cot \theta +\frac{2\Lambda
_{,\theta }}{r^{2}\Lambda ^{3}}=0,  \label{4}
\end{equation}%
\begin{equation}
G_{\theta }^{\theta }=\frac{f\Lambda _{,r}^{2}}{\Lambda ^{4}}-\frac{%
2f\Lambda _{,r}}{r\Lambda ^{3}}+\frac{f^{\prime }}{r\Lambda ^{2}}+\frac{%
\Lambda _{,\theta }^{2}}{r^{2}\Lambda ^{4}}-\frac{2\Lambda _{,\theta }}{%
r^{2}\Lambda ^{3}}\cot \theta +\frac{f^{\prime \prime }}{2\Lambda ^{2}}=0,
\label{5}
\end{equation}%
and%
\begin{equation}
G_{\phi }^{\phi }=\frac{2f\Lambda _{,r}}{r\Lambda ^{3}}+\frac{2f^{\prime
}\Lambda _{,r}}{\Lambda ^{3}}+\frac{2f\Lambda _{,rr}}{\Lambda ^{3}}+\frac{%
2\Lambda _{,\theta \theta }}{r^{2}\Lambda ^{3}}-\frac{\Lambda _{,\theta }^{2}%
}{r^{2}\Lambda ^{4}}-\frac{f\Lambda _{,r}^{2}}{\Lambda ^{4}}+\frac{f^{\prime
}}{r\Lambda ^{2}}+\frac{f^{\prime \prime }}{2\Lambda ^{2}}=0,  \label{6}
\end{equation}%
where a prime stands for the derivative with respect to $r$ and $\Lambda
_{,x}=\frac{\partial \Lambda }{\partial x}.$ Combining (\ref{2}) and (\ref{3}%
) yields%
\begin{equation}
\Lambda _{,r}\left( r\Lambda _{,r}-2\Lambda \right) =0,  \label{7}
\end{equation}%
which admits two solutions, given by%
\begin{equation}
\Lambda \left( r,\theta \right) =\lambda \left( \theta \right) ,  \label{8}
\end{equation}%
and%
\begin{equation}
\Lambda \left( r,\theta \right) =\lambda \left( \theta \right) r^{2},
\label{9}
\end{equation}%
in which $\lambda \left( \theta \right) $ is an arbitrary function of $%
\theta .$ From (\ref{4}), we observe that the first solution i.e., Eq. (\ref%
{8}) yields the trivial solution $\lambda \left( \theta \right) =const.$
which leads to the Schwarzschild black hole. On the other hand $\Lambda
\left( r,\theta \right) =\lambda \left( \theta \right) r^{2}$ is the
non-trivial solution which after introducing into (\ref{4}) implies%
\begin{equation}
\lambda ^{\prime }\left( \theta \right) -2\cot \theta \lambda \left( \theta
\right) =0.  \label{10}
\end{equation}%
The latter equation is solved to obtain the exact form of $\lambda \left(
\theta \right) $ which is given by 
\begin{equation}
\lambda \left( \theta \right) =\lambda _{0}\sin ^{2}\theta ,  \label{11}
\end{equation}%
where $\lambda _{0}$ is an integration constant. As it will be shown in the
sequel, $\lambda _{0}$\ is related to the linear mass density of the
solution. Upon considering (\ref{11}) and (\ref{9}) in the field equations (%
\ref{2})-(\ref{6}) they reduce to the following equations%
\begin{equation}
G_{t}^{t}=G_{r}^{r}=\frac{rf^{\prime }+f-1}{\lambda _{0}^{2}r^{6}\sin
^{4}\theta }=0,  \label{12}
\end{equation}%
\begin{equation}
G_{\theta }^{\theta }=\frac{rf^{\prime \prime }+2f^{\prime }}{2\lambda
_{0}^{2}r^{5}\sin ^{4}\theta }=0,  \label{13}
\end{equation}%
and%
\begin{equation}
G_{\phi }^{\phi }=\frac{r^{2}f^{\prime \prime }+10rf^{\prime }+8f-8}{\lambda
_{0}^{2}r^{6}\sin ^{4}\theta }=0.  \label{14}
\end{equation}%
Upon solving (\ref{12}) which satisfies the other two equations too, the
solution for $f\left( r\right) $ is found to be%
\begin{equation}
f\left( r\right) =1-\frac{2m}{r},  \label{15}
\end{equation}%
where $m$ is another integration constant. Finally, in the spherically
symmetric coordinates i.e., $\left\{ t,r,\theta ,\phi \right\} ,$ the black
hole solution is expressed by the line element 
\begin{equation}
ds^{2}=\lambda _{0}^{2}r^{4}\sin ^{4}\theta \left( -\left( 1-\frac{2m}{r}%
\right) dt^{2}+\frac{dr^{2}}{1-\frac{2m}{r}}+r^{2}d\theta ^{2}\right) +\frac{%
d\phi ^{2}}{\lambda _{0}^{2}r^{2}\sin ^{2}\theta }.  \label{16}
\end{equation}%
This is the solution of Einstein's vacuum field equations i.e., $G_{\mu \nu
}=0,$ and since the Ricci scalar is also zero they are the solution of $%
R_{\mu \nu }=0$, too. Although the Ricci scalar and the Ricci tensors are
zero the Kretschmann scalar is non-zero and is given by%
\begin{equation}
\mathcal{K}=\frac{1008m^{2}}{\lambda _{0}^{4}r^{14}\sin ^{8}\theta }-\frac{%
288\left( 2+\sin ^{2}\theta \right) m}{\lambda _{0}^{4}r^{13}\sin
^{10}\theta }+\frac{192}{\lambda _{0}^{4}r^{12}\sin ^{12}\theta }.
\label{17}
\end{equation}%
For convenience one may express (\ref{16}) in the prolate coordinates i.e., $%
\left\{ T,X,Y,\phi \right\} $ where $T=\sqrt{\lambda _{0}}t,$ $X=\sqrt{%
\lambda _{0}}r-1$ and $Y=\cos \theta $. Hence, (\ref{16}) reads%
\begin{equation}
d\tilde{s}^{2}=\lambda _{0}ds^{2}=\left( X+1\right) ^{4}\left(
1-Y^{2}\right) ^{2}\left( -\left( 1-\frac{2\tilde{m}}{X+1}\right) dT^{2}+%
\frac{dX^{2}}{1-\frac{2\tilde{m}}{X+1}}+\frac{\left( X+1\right) ^{2}}{1-Y^{2}%
}dY^{2}\right) +\frac{d\phi ^{2}}{\left( X+1\right) ^{2}\left(
1-Y^{2}\right) },  \label{18}
\end{equation}%
in which $\tilde{m}=\sqrt{\lambda _{0}}m.$ We add that while $r\in \left[
0,\infty \right) $ and $\theta \in \left[ 0,\pi \right] $, the prolate
coordinates are defined such that $X\in \left[ -1,\infty \right) $ and $Y\in %
\left[ -1,1\right] .$ In terms of the prolate coordinates the Kretschmann
scalar is given by%
\begin{equation}
\mathcal{K}=\frac{1008\tilde{m}^{2}\lambda _{0}^{2}}{\left( X+1\right)
^{14}\left( 1-Y^{2}\right) ^{4}}-\frac{288\left( 3-Y^{2}\right) \tilde{m}%
\lambda _{0}^{2}}{\left( X+1\right) ^{13}\left( 1-Y^{2}\right) ^{5}}+\frac{%
192\lambda _{0}^{2}}{\left( X+1\right) ^{12}\left( 1-Y^{2}\right) ^{6}}.
\label{19}
\end{equation}%
From (\ref{17}) and (\ref{19}), we observe that the axis of symmetry i.e.,
the line $\theta =0,\pi /Y=\pm 1$ is the singularity of the black hole. Such
singularity is due to the mass distribution on the symmetry axis. With $m=0$
the solution (\ref{16}) can be expressed in cylindrical coordinates i.e., $%
\left\{ t,\rho ,z,\phi \right\} $ where $\rho =r\sin \theta $ and $z=r\cos
\theta .$ In such setting (\ref{16}) reads as 
\begin{equation}
ds^{2}=\lambda _{0}^{2}\rho ^{4}\left( -dt^{2}+d\rho ^{2}+dz^{2}\right) +%
\frac{d\phi ^{2}}{\lambda _{0}^{2}\rho ^{2}}.  \label{20}
\end{equation}%
Remarkably, (\ref{20}) is the well-known Levi-Civita spacetime with the
general line element given by%
\begin{equation}
ds^{2}=\rho ^{2\sigma \left( \sigma -1\right) }\left( -dt^{2}+d\rho
^{2}\right) +\rho ^{2\sigma }dz^{2}+\alpha ^{2}\rho ^{2\left( 1-\sigma
\right) }d\phi ^{2},  \label{21}
\end{equation}%
where the Levi-Civita constant is $\sigma =2$. Therefore, the black hole
solution (\ref{16}) may be called the Schwarzschild-Levi-Civita black hole
as we mentioned before. We should add that since the singularity extends to
infinity on the axis of symmetry the mass is not finite and therefore an ADM
mass is not defined. Hence, $m$\ is not the mass of the black hole. Let us
express the solution in cylindrical coordinates to obtain a reasonable mass
density. Introducing the transformation%
\begin{equation}
\rho =rH\left( r\right) \sin \theta ,  \label{R18}
\end{equation}%
and%
\begin{equation}
z=rH\left( r\right) \cos \theta ,  \label{R19}
\end{equation}%
where 
\begin{equation}
\frac{d}{dr}\ln \left( rH\left( r\right) \right) =\frac{1}{r\sqrt{1-\frac{2m%
}{r}}},  \label{R20}
\end{equation}%
the black hole solution transforms into%
\begin{multline}
ds^{2}=-\frac{\lambda _{0}^{2}\rho ^{4}}{2^{4}}\left( 1+\frac{m}{\sqrt{\rho
^{2}+z^{2}}}\right) ^{4}\left( 1-\frac{m^{2}}{\rho ^{2}+z^{2}}\right)
^{2}dt^{2}+  \label{R29} \\
\frac{\lambda _{0}^{2}\rho ^{4}}{2^{6}}\left( 1+\frac{m}{\sqrt{\rho
^{2}+z^{2}}}\right) ^{12}\left( d\rho ^{2}+dz^{2}\right) +\frac{4}{\lambda
_{0}^{2}\rho ^{2}\left( 1+\frac{m}{\sqrt{\rho ^{2}+z^{2}}}\right) ^{4}}d\phi
^{2}.
\end{multline}%
In comparison with the Weyl-Lewis-Papapetrou class of static and axially
symmetric solution 
\begin{equation}
ds^{2}=-e^{2\upsilon }dt^{2}+e^{2\mu }\left( d\rho ^{2}+dz^{2}\right)
+B^{2}e^{-2\upsilon }\rho ^{2}d\phi ^{2},  \label{R22}
\end{equation}%
in which $\mu ,\nu $\ and $B$\ are functions of $\rho $\ and $z.$, one finds%
\begin{equation}
e^{2\upsilon }=\frac{\lambda _{0}^{2}\rho ^{4}}{2^{4}}\left( 1+\frac{m}{%
\sqrt{\rho ^{2}+z^{2}}}\right) ^{4}\left( 1-\frac{m^{2}}{\rho ^{2}+z^{2}}%
\right) ^{2},  \label{R23}
\end{equation}%
\begin{equation}
e^{2\mu }=\frac{\rho ^{4}}{2^{6}}\left( 1+\frac{m}{\sqrt{\rho ^{2}+z^{2}}}%
\right) ^{12},  \label{R24}
\end{equation}%
and%
\begin{equation}
B=\frac{1}{2}\left( 1-\frac{m^{2}}{\rho ^{2}+z^{2}}\right) .  \label{R25}
\end{equation}%
Similar to the Weyl class of solutions, the Killing vector fields of the
black hole (\ref{R29}) are $\partial _{t}$\ and $\partial _{\phi }.$\
Furthermore, (\ref{R29}) in the limit $\rho \rightarrow \infty ,$\ upon
rescaling $\rho \rightarrow 2\rho ,$\ $z\rightarrow 2z$\ and $t\rightarrow 
\frac{1}{\lambda _{0}}t$\ yields%
\begin{equation}
ds^{2}=-\rho ^{4}dt^{2}+\rho ^{4}\left( d\rho ^{2}+dz^{2}\right) +\frac{1}{%
\lambda _{0}^{2}}\rho ^{-2}d\phi ^{2},  \label{R26}
\end{equation}%
which is the Levi-Civita metric in Weyl coordinates with $\sigma =2$. As the
ADM mass of the Levi-Civita metric is not defined, we introduce the Komar
linear mass density which is given by $\mu _{K}=\frac{1}{\lambda _{0}}$. For
the detail of calculating the Komar linear mass density of the Levi-Civita
metric, we refer to \cite{Komar}. On the other hand since the asymptotic
behavior of the black hole solution (\ref{R29}) is the Levi-Civita metric
they naturally admit the same Komar mass density. Therefore the black hole
solution (\ref{R29}) is powered by a linear mass on its axis of symmetry
namely the $z-$axis with a Komar mass linear density equal to its asymptotic
Levi-Civita solution i.e., $\mu _{K}=\frac{1}{\lambda _{0}}.$

Finally, due to the asymptotic behavior of the black hole solution (\ref{16}%
) that is the Levi-Civita spacetime, we interpret the line element (\ref{16}%
) to be the Schwarzschild black hole formed in the Levi-Civita spacetime
ambiance.

\section{The photon orbit}

In this part, we investigate the possible circular geodesic of a null
particle. To do so, we start with the effective Lagrangian density of a null
particle moving in the vicinity of the black hole (\ref{16}) that is given by%
\begin{equation}
2\mathcal{L}=r^{4}\sin ^{4}\theta \left( -\left( 1-\frac{r_{+}}{r}\right) 
\dot{t}^{2}+\frac{\dot{r}^{2}}{1-\frac{r_{+}}{r}}+r^{2}\dot{\theta}%
^{2}\right) +\frac{\dot{\phi}^{2}}{r^{2}\sin ^{2}\theta },  \label{23}
\end{equation}%
in which an over-dot stands for the derivative with respect to an affine
parameter and $r_{+}=2m$ is the event horizon of the black hole. The
conserved energy and angular momentum are given by%
\begin{equation}
E=-\frac{\partial \mathcal{L}}{\partial \dot{t}}=r^{4}\sin ^{4}\theta \left(
1-\frac{r_{+}}{r}\right) \dot{t},  \label{24}
\end{equation}%
and%
\begin{equation}
\ell =\frac{\partial \mathcal{L}}{\partial \dot{\phi}}=\frac{\dot{\phi}}{%
r^{2}\sin ^{2}\theta }.  \label{25}
\end{equation}%
The $\theta $-component of the geodesic equation is not trivially satisfied
by a constant $\theta,$ however, it is satisfied by $\theta =\frac{\pi }{2}$%
. Therefore, for the rest of this section, we set $\theta =\frac{\pi }{2}$
which corresponds to the geodesic of the null particle on the equatorial
plane. The main geodesic equation is obtained by applying the constraint $%
\dot{x}_{\mu }\dot{x}^{\mu }=0$ for the null geodesic which implies%
\begin{equation}
r^{8}\dot{r}^{2}+r^{6}\ell ^{2}\left( 1-\frac{r_{+}}{r}\right) =E^{2}.
\label{26}
\end{equation}%
For the circular orbit we set $r=r_{c}$ and consequently $\dot{r}=\ddot{r}=0$
such that considering (\ref{26}) we find%
\begin{equation}
r_{c}^{5}\left( r_{c}-r_{+}\right) =\frac{E^{2}}{\ell ^{2}},  \label{27}
\end{equation}%
and 
\begin{equation}
r_{c}=\frac{5}{6}r_{+}.  \label{28}
\end{equation}%
The latter shows that $r_{c}<r_{+}$ and therefore it is inside the black
hole. In other words, there is no circular orbit outside the black hole
which is unlike the Schwarzschild black hole whose photonsphere located at $%
r_{c}=\frac{3}{2}r_{+}.$ This simple geodesic implies that the Schwarzschild
black hole formed in the Levi-Civita spacetime possesses different
fundamental structure than the standard Schwarzschild black hole.

\section{Petrov Type classification}

Starting from the complex null tetrad basis one-form in the Newman-Penrose
(NP) formalism 
\begin{equation}
\sqrt{2}\mathbf{l}=\lambda _{0}r^{2}\sin ^{2}\theta \sqrt{1-\frac{2m}{r}}dt-%
\frac{1}{\lambda _{0}r\sin \theta }d\phi ,  \label{R1}
\end{equation}%
\begin{equation}
\sqrt{2}\mathbf{n}=\lambda _{0}r^{2}\sin ^{2}\theta \sqrt{1-\frac{2m}{r}}dt+%
\frac{1}{\lambda _{0}r\sin \theta }d\phi ,  \label{R2}
\end{equation}%
\begin{equation}
\sqrt{2}\mathbf{m}=\frac{\lambda _{0}r^{2}\sin ^{2}\theta }{\sqrt{1-\frac{2m%
}{r}}}dr+i\lambda _{0}r^{3}\sin ^{2}\theta d\theta ,  \label{R3}
\end{equation}%
\begin{equation}
\sqrt{2}\mathbf{m}^{\ast }=\frac{\lambda _{0}r^{2}\sin ^{2}\theta }{\sqrt{1-%
\frac{2m}{r}}}dr-i\lambda _{0}r^{3}\sin ^{2}\theta d\theta  \label{R3s}
\end{equation}%
and the nonzero Weyl tensor components%
\begin{equation}
C_{trtr}=-6\lambda _{0}^{2}r\sin ^{2}\theta \left( \cos ^{2}\theta \left(
m-r\right) -m+\frac{r}{3}\right) ,  \label{R4}
\end{equation}%
\begin{equation}
C_{trt\theta }=-6\lambda _{0}^{2}r^{3}\sin ^{3}\theta \cos \theta \left( 1-%
\frac{2m}{r}\right) ,  \label{R5}
\end{equation}%
\begin{equation}
C_{t\theta t\theta }=\lambda _{0}^{2}r^{3}\sin ^{2}\theta \left( 1-\frac{2m}{%
r}\right) \left( \cos ^{2}\theta \left( 9m-6r\right) -9m+4r\right) ,
\label{R6}
\end{equation}%
\begin{equation}
C_{t\phi t\phi }=-\left( 1-\frac{2m}{r}\right) \frac{\left( 2r-3m\sin
^{2}\theta \right) }{\lambda _{0}^{2}r^{5}\sin ^{4}\theta },  \label{R7}
\end{equation}%
\begin{equation}
C_{r\theta r\theta }=\frac{\lambda _{0}^{2}r^{3}\sin ^{4}\theta \left(
2r-3m\sin ^{2}\theta \right) }{\left( 1-\frac{2m}{r}\right) },  \label{R8}
\end{equation}%
\begin{equation}
C_{r\phi r\phi }=-\frac{\left( \cos ^{2}\theta \left( 9m-6r\right)
-9m+4r\right) }{\lambda _{0}^{2}r^{5}\sin ^{4}\theta \left( 1-\frac{2m}{r}%
\right) },  \label{R9}
\end{equation}%
\begin{equation}
C_{r\phi \theta \phi }=-\frac{6\cos \theta }{\lambda _{0}^{2}r^{3}\sin
^{3}\theta },  \label{R10}
\end{equation}%
\begin{equation}
C_{\theta \phi \theta \phi }=\frac{2\left( 3\cos ^{2}\theta \left(
m-r\right) -3m+r\right) }{\lambda _{0}^{2}r^{3}\sin ^{4}\theta },
\label{R11}
\end{equation}%
we calculate the Weyl-NP scalars $\Psi _{i}$\ which are expressed by%
\begin{equation}
\Psi _{0}=C_{\alpha \beta \gamma \delta }l^{\alpha }m^{\beta }l^{\gamma
}m^{\delta }=\frac{-3\left( 1-\frac{2m}{r}\right) }{\lambda
_{0}^{2}r^{3}\sin ^{4}\theta },  \label{R12}
\end{equation}%
\begin{equation}
\Psi _{1}=C_{\alpha \beta \gamma \delta }l^{\alpha }n^{\beta }l^{\gamma
}m^{\delta }=\frac{3\cos \theta \sqrt{1-\frac{2m}{r}}}{\lambda
_{0}^{2}r^{6}\sin ^{5}\theta },  \label{R13}
\end{equation}%
\begin{equation}
\Psi _{2}=C_{\alpha \beta \gamma \delta }l^{\alpha }m^{\beta }l^{\gamma
}m^{\ast \delta }=-\frac{3\left( 1-\left( 1-\frac{m}{r}\right) \sin
^{2}\theta \right) }{\lambda _{0}^{2}r^{6}\sin ^{6}\theta },  \label{R14}
\end{equation}%
\begin{equation}
\Psi _{3}=C_{\alpha \beta \gamma \delta }l^{\alpha }n^{\beta }m^{\ast \gamma
}n^{\delta }=\frac{-3\cos \theta \sqrt{1-\frac{2m}{r}}}{\lambda
_{0}^{2}r^{12}\sin ^{5}\theta },  \label{R15}
\end{equation}%
and%
\begin{equation}
\Psi _{4}=C_{\alpha \beta \gamma \delta }n^{\alpha }m^{\ast \beta }n^{\gamma
}m^{\ast \delta }=\frac{-3\left( 1-\frac{2m}{r}\right) }{\lambda
_{0}^{2}r^{6}\sin ^{4}\theta }.  \label{R16}
\end{equation}%
Next, the Petrov type of the solution is determined by the number of the
distinct roots $z$\ of the following equation%
\begin{equation}
\Psi _{0}+4z\Psi _{1}+6z^{2}\Psi _{2}+4z^{3}\Psi _{3}+z^{4}\Psi _{4}=0.
\label{R17}
\end{equation}%
Having all Weyl-NP scalars $\Psi _{i}$\ nonzero, our detailed calculation
shows that there are 4 distinct roots for the latter equation, indicating
that the spacetime is algebraically general, Petrov type I. This is
remarkable to know that both the Schwarzschild black hole and the
Levi-Civita spacetime are of Petrov type D which can be found in different
limits of the Plebanski-Demianski (PD) 7-parameters class of solutions \cite%
{PD1,PD2}. This implies that the Schwarzschild-Levi-Civita black hole is a
new solution that departed from the PD class of solutions. Finally, the
Weyl-NP scalars once more imply the occurrence of the singularity at the
axis of symmetry with $\theta =0,\pi $, and $r=0$.

\section{Conclusion}

In this research, we introduced a new exact solution to Einstein's field
equation in vacuum. The idea of the existence of such a solution was taken
from the Ernst black hole which is an interpolation between the
Schwarzschild black hole and the Melvin universe. Our exact solution is
effectively a two-parameter black hole that interpolates the Schwarzschild
black hole and the Levi-Civita spacetime such that we called it the
Schwarzschild-Levi-Civita black hole. This has to be added that while the
new solution is asymptotically the Levi-Civita spacetime, it doesn't admit
the Schwarzschild black hole solution in any limit. It is more like the
Schwarzschild black hole is formed in the Levi-Civita spacetime. This is in
analogy to the Ernst solution where a Schwarzschild black hole is formed in
the Melvin universe. It is static and axially symmetric with the singularity
at its axis of symmetry. Unlike the Ernst black hole which is the solution
of the Einstein-Maxwell field equations, the Schwarzschild-Levi-Civita black
hole is the solution of the Einstein equations in vacuum. We have studied
the existence of the photonsphere and shown that unlike in the Schwarzschild
black hole, there exists no photonsphere outside the black hole. This is an
indication that this black hole is a non-trivial extension of the
Schwarzschild black hole. Further investigation on the physical properties
of the new solution is open.

\end{document}